\documentclass[cits,natbib]{PoS}

\def \inte {INTEGRAL}

\def \ATel {The Astronomer's Telegram}
\newcommand\sw{{\it Swift}}

\newcommand\apj{{ApJ}}%
\newcommand\apjl{{ApJ}}%
\newcommand\aap{{A\&A}}%
\newcommand\mnras{{MNRAS}}%
\title{The \emph{Swift} Supergiant Fast X-ray Transients Project: recent results}

\ShortTitle{SFXTs with {\it Swift}}

\author{ \speaker{P.\ Romano},$^a$ S.\ Vercellone,$^a$ V. La Parola,$^a$ G.\ Cusumano,$^a$ V.\ Mangano,$^a$ 
P.~Esposito,$^b$ J.A.\ Kennea,$^c$ D.N.\ Burrows,$^c$ H.A.\ Krimm,$^{de}$ C.\ Pagani,$^f$N.\ Gehrels,$^g$ \\
\llap{$^a$}INAF, Istituto di Astrofisica Spaziale e Fisica Cosmica, \\
         Via U.\ La Malfa 153, I-90146 Palermo, Italy \\
\llap{$^b$}INAF, Osservatorio Astronomico di Cagliari, \\
         localit\`a Poggio dei Pini, strada 54, I-09012 Capoterra, Italy\\
\llap{$^c$}Department of Astronomy and Astrophysics, Pennsylvania State  University, \\
         University Park, PA 16802, USA\\
\llap{$^d$}CRESST/Goddard Space Flight Center, Greenbelt, MD, USA\\
\llap{$^e$}Universities Space Research Association, Columbia, MD, USA\\
\llap{$^f$}Department of Physics \& Astronomy, University of Leicester, LE1 7RH, UK\\
\llap{$^g$}NASA/Goddard Space Flight Center, Greenbelt, MD 20771, USA\\
E-mail: \email{romano@ifc.inaf.it}  \hspace{1cm} \\   \href{http://www.ifc.inaf.it/sfxt/}{http://www.ifc.inaf.it/sfxt/ }
}

\abstract{
We present an overview of our Supergiant Fast X-ray Transients (SFXT) project, 
that started in 2007, by highlighting the unique observational contribution 
\sw{} is giving to this exciting new field. 
By means of outburst detection with \sw/BAT and follow-up with \sw/XRT,  
we demonstrated that while the brightest phase of the outburst only lasts a few hours, 
further significant activity is observed at lower fluxes for a considerably 
longer (weeks) time. 
After intense monitoring with \sw/XRT, we now have a firm estimate of the time SFXTs
spend in each phase. The 4 SFXTs we monitored for 1--2 years spend between 3 and 5\,\% 
of the time in bright outbursts. The most most probable flux level at which a random 
observation will find these sources, when detected, is 
$F_{\rm 2-10\,keV}\sim 1$--$2\times10^{-11}$ erg cm$^{-2}$ s$^{-1}$ (unabsorbed), 
corresponding to luminosities of a few $10^{33}$ to a few $10^{34}$ erg s$^{-1}$.
Finally,  the duty-cycle of inactivity ranges between 19 and 55\,\%.
}

\FullConference{25th Texas Symposium on Relativistic Astrophysics - TEXAS 2010\\
		December 06-10, 2010\\
		Heidelberg, Germany}

\begin{document}

\section{Introduction}

Supergiant fast X--ray transients (SFXTs) 
are a recently discovered class of transient High-Mass X--ray Binaries 
(e.g.\ \cite{Sguera2005}) probably hosting a neutron star accretor, 
whose optical counterparts are blue supergiant stars. 
Their X--ray outbursts, that last a few hours in the hard X--ray band, 
are about 4 orders of magnitude brighter than their quiescent state 
(peak luminosities of $L\sim 10^{36}$--$10^{37}$\,erg\,s$^{-1}$ 
\cite{Sguera2005,Negueruela2006:ESASP604},  
versus quiescent $L \sim 10^{32}$\,erg\,s$^{-1}$, 
e.g.\ \cite{zand2005,Bozzo2010:quiesc1739n08408}). The physics of 
the bright outbursts is still being debated,  
and it is probably related to either the properties of 
the wind from the supergiant companion 
\cite{zand2005,Walter2007,Negueruela2008,Sidoli2007} or to the 
presence of a centrifugal or magnetic barrier \cite{Grebenev2007,Bozzo2008}. 
In this paper we shall present an overview of our SFXT project\footnote{http://www.ifc.inaf.it/sfxt/ }, 
that started about four years ago, and thus highlight 
the unique observational contribution \sw{} \cite{Gehrels2004} 
is giving to this exciting new field.

\section{The {\emph Swift} contribution}

Why \sw? 
\sw{} has unique characteristics that make it particularly well-suited to 
study both the SFXT bright outbursts and the fainter out-of-outburst states.   
Indeed, it combines a fast-slewing capability and a broad-band energy coverage, 
so that it can easily catch outbursts from these fast transients in their 
very early stages and study them panchromatically as they evolve in time. 
These are paired with a unique flexible observing scheduling,
which allows monitoring efforts that cover all phases of their lives 
with a high sensitivity in the soft X-ray regime. 

\begin{figure}[b]
 \includegraphics[height=0.65\textheight,width=0.3\textwidth,angle=270]{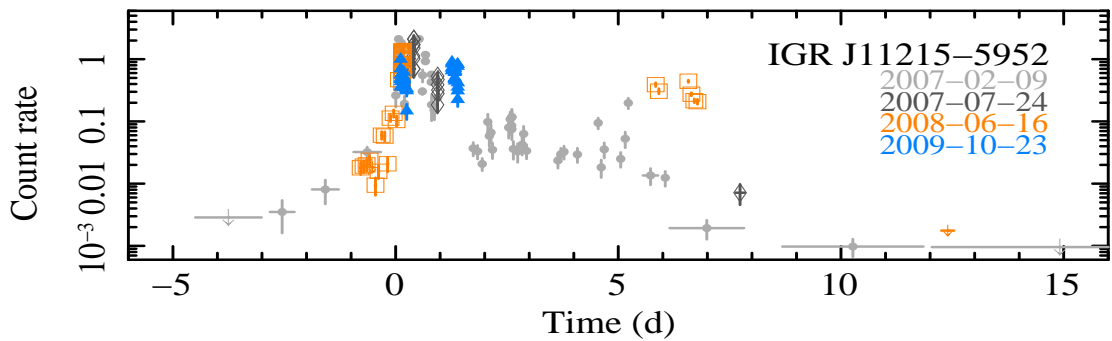}
\caption{\sw/XRT light curves of the  2007 Feb 9 outbursts (light grey filled circles), 
	        2007 Jul 24 (dark grey empty diamonds), 2008 Jun 16 (orange empty squares), 
                and 2009 Nov 23 (cyan filled triangles), folded with a period of 164.6 days,
                relative to the peak of the first outburst.  
Data from \cite{Romano2007,Sidoli2007,Romano2009:11215_2008,Esposito2009:atel2257}. 
}
\label{texas10:fig:11215}
\end{figure}

Before 2007 most observations of SFXT outbursts were fortuitous (e.g.\ \cite{zand2005}). 
At the time, IGR~J11215$-$5952 was the only SFXT with periodic outbursts \cite{SidoliPM2006}, and  
this gave us an unprecedented opportunity to actually plan to observe the first available outburst,
that was predicted to occur on 2007 Feb 9. 
We performed a target of opportunity (TOO) \sw/X--ray Telescope (XRT, \cite{Burrows2005:XRT}) 
monitoring campaign  \cite{Romano2007} and could follow the event for 23 days and 
for three orders of magnitude in flux, from non-detection up to the peak of the outburst at 
$10^{36}$ erg\,s$^{-1}$, and back down until it disappeared again below our detection limits
(Figure~\ref{texas10:fig:11215}). 
The surprising discovery was that while the bright outburst does last only a few hours, 
further significant activity (hence accretion onto the compact object) 
is seen at lower fluxes for a considerably longer (weeks) time. 
A further \sw{} contribution is the determination of the orbital period of $\sim 164.6$\,d
 \cite{Sidoli2007,Romano2009:11215_2008} (the INTEGRAL period was 330\,d, \cite{SidoliPM2006}),
a rare instance, as orbital periods are generally found from all-sky monitor data.

Since 2007, \sw{} has devoted considerable observing time to perform a systematic
study of SFXTs, with a strategy that combined 
monitoring programs with outburst follow-up observations. 
Our sample for long term monitoring consists of 4 targets, chosen among the 8 
SFXTs known at the time and it includes, as ordered in Table~\ref{texas10:tab:campaign},  
one object that triggered the \sw/Burst Alert Telescope (BAT; \cite{Barthelmy2005:BAT})
previously, the two prototypes of the class, and a pulsar.  
Starting on 2007 Oct we observed them two to three times a week for 1--2\,ks each.
Table~\ref{texas10:tab:campaign} reports observing date ranges (col.\,2), number of 
observations (col.\,3), and total on-source times (col.\,4). 
For this project, the BAT on-board catalog was modified to allow the BAT to 
trigger on all known and candidate SFXTs as if they were GRBs, ensuring rapid 
followup of outbursts detected by the BAT.
This strategy proved to be an efficient way to catch outbursts with BAT 
and to monitor them with XRT during their evolution 
\cite{Romano2008:sfxts_paperII,Romano2009:sfxts_paper08408,Sidoli2009:sfxts_paperIII,
Sidoli2009:sfxts_paperIV,Sidoli2009:sfxts_sax1818,Romano2009:sfxts_paperV,Romano2011:sfxts_paperVI,Romano2011:sfxts_paperVII}. 
The regular monitoring also allowed us, for the very first time, to study the long term properties 
of this class of objects with a highly sensitive instrument, in particular, the lowest states 
towards quiescence \cite{Sidoli2008:sfxts_paperI,Romano2009:sfxts_paperV,Romano2011:sfxts_paperVI}. 

\begin{figure}[t]
		\vspace{-3truecm}
 \includegraphics[width=0.9\textwidth,height=0.7\textheight,angle=0]{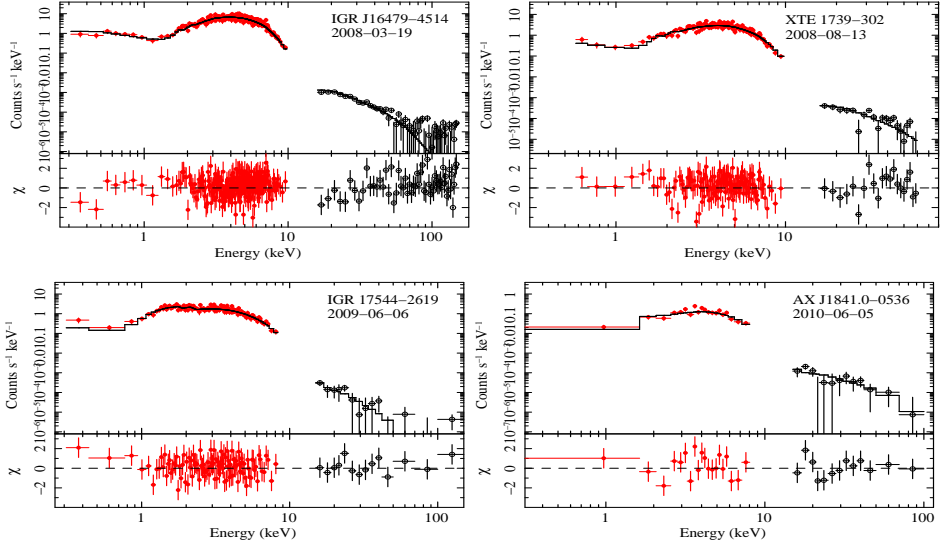}
		\vspace{-5.5truecm}
\caption{Spectroscopy of one representative outburst of each of the four SFXTs monitored by \sw. 
Filled (red) circles and empty (black) circles denote XRT and BAT data, respectively. 
The data were fit with an absorbed power-law model with a high energy cut-off ({\tt highecut} in XSPEC). 
Data from \cite{Romano2008:sfxts_paperII,Sidoli2009:sfxts_paperIII,Romano2011:sfxts_paperVI,Romano2011:sfxts_paperVII}. 
}
\label{texas10:fig:outburst_spectra}
\end{figure}

Simultaneous observations with XRT and BAT allowed us to perform 
broad band spectroscopy of SFXT outbursts for the first time \cite{Romano2008:sfxts_paperII}, 
and this, we note, is the true strength of \sw. 
This is particularly important because of the general shape of the SFXT spectrum
in outburst, a hard power law below 10\,keV with an exponential cutoff at 15--30\,keV. 
With the large (0.3--150\,keV) \sw{} energy range we can both constrain the 
hard-X spectral properties (to compare with popular 
accreting neutron star models) and obtain a handle on the absorption. 
As an example, Figure~\ref{texas10:fig:outburst_spectra} shows the spectrum of one outburst of each 
of the four SFXTs monitored, while Table~\ref{texas10:tab:outburst_spectra} reports the 
values of the spectral parameters.

\begin{table}[t]
\begin{tabular}{lccccccc}
\hline
\hline
Name             &Campaign Dates                &Obs.\     &Expo. &$\Delta T_{\Sigma}$  & $P_{\rm short}$ &  IDC  & Rate$_{\Delta T_{\Sigma}}$   \\
    		                                       &  &N. & (ks) &(ks) & (\%) &  (\%) &  	\\ 	
\hline
IGR~J16479--4514 &  2007-10-26--2009-11-01	& 144&	161 & 29.7 &3  & 19 & $3.1\pm0.5$ \\	      
XTE~J1739--302 	 &  2007-10-27--2009-11-01	& 184&	206 & 71.5 &10 & 39 & $4.0\pm0.3$ \\
IGR~J17544--2619 &  2007-10-28--2009-11-03	& 142&	143 & 69.3 &10 & 55 & $2.2\pm0.2$ \\
AX~J1841.0--0536 &  2007-10-26--2008-11-15	&  88&	 96 & 26.6 &3  & 28 & $2.4\pm0.4$ \\  
 Total           &                             & 558&606 & & & & \\
\hline
\end{tabular}
\caption{The \sw{} long-term monitoring campaign. 
   $\Delta T_{\Sigma}$ is sum of the exposures accumulated in all observations (with exposure $>900$\,s)  
   where only a 3-$\sigma$ upper limit was achieved;  
   $P_{\rm short}$ is the percentage of time lost to short observations; 
   IDC is the duty cycle of inactivity, i.e.,  
   the time each source spends undetected down to a flux limit of 1--3$\times10^{-12}$ erg cm$^{-2}$ s$^{-1}$;
   Rate$_{\Delta T_{\Sigma}}$ is the cumulative count rate (0.2--10\,keV, $\times10^{-3}$ counts s$^{-1}$). 
   Adapted from \cite{Romano2009:sfxts_paperV,Romano2011:sfxts_paperVI}.  
\label{texas10:tab:campaign} }
\end{table}

\begin{table}[t]
\begin{tabular}{lcccccc}
\hline
\hline
Name             &Date         & $N_{\rm H}$ & $\Gamma$ & $E_{\rm cut}$ & $E_{\rm fold}$ & Reference\\           
                 & &  ($10^{22}$~cm$^{-2}$) &          & (keV)       & (keV)         &     \\  
\hline
IGR~J16479$-$4514 &2008-03-19 & $6.2_{-0.5}^{+0.6}$ &$1.2_{-0.1}^{+0.2}$ &$6_{-1}^{+1}$ &$15_{-2}^{+3}$ & \cite{Romano2008:sfxts_paperII} \\
XTE~J1739$-$302   &2008-08-13 &$4.8_{-0.6}^{+1.3}$ &$0.8_{-0.2}^{+0.4}$   &$5_{-1}^{+2}$       &$12_{-3}^{+8}$      & This work \\
IGR~J17544$-$2619 &2009-06-06 &$1.0_{-0.3}^{+0.2}$  &$0.6_{-0.4}^{+0.2}$   &$3_{-1}^{+1}$ &$8_{-3}^{+4}$ & \cite{Romano2011:sfxts_paperVI} \\
AX~J1841.0$-$0536 &2010-06-05 &$1.9_{-1.0}^{+1.7}$ &$0.2_{-0.5}^{+0.4}$ &$4_{-4}^{+12}$ &$16_{-9}^{+10}$ & \cite{Romano2011:sfxts_paperVII}  \\
\hline
\end{tabular}
\caption{Spectral parameters of the outbursts shown in Figure~2. 
}
\label{texas10:tab:outburst_spectra}
\end{table}

\begin{figure}[H]
                \vspace{-1.1truecm}
                \hspace{-1truecm}
 \includegraphics[height=0.8\textheight,angle=90]{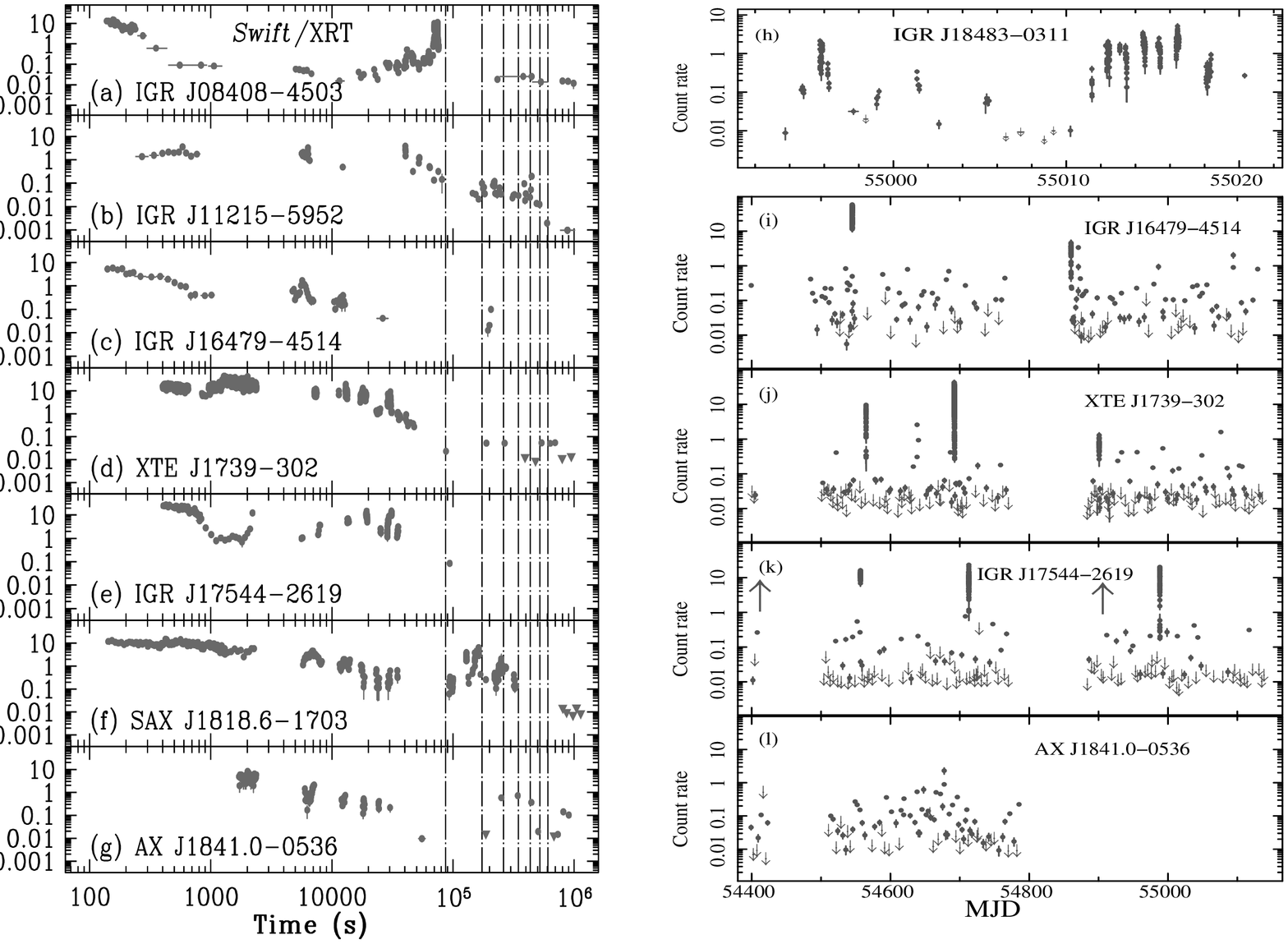}
                \vspace{-1.5truecm}
\caption[XRT light curves]{
{\bf Left:} 
\sw/XRT light curves of the most representative {\bf outbursts} of SFXTs 
followed by \sw/XRT,  
referred to their respective BAT triggers
(IGR~J11215$-$5952 did not trigger the BAT, so it is referred to MJD 54139.94).
Points denote detections (binning of at least 20 counts bin$^{-1}$), triangles 3$\sigma$ upper limits.  
Vertical dashed lines mark time intervals equal to 1 day, up to a week. 
References: 
(a) IGR~J08408--4503   (2008-07-05, \cite{Romano2009:sfxts_paper08408});
(b) IGR~J11215$-$5952  (2007-02-09, \cite{Romano2007}); 
(c) IGR~J16479$-$4514  (2005-08-30, \cite{Sidoli2008:sfxts_paperI}); 
(d) XTE~J1739$-$302    (2008-08-13, \cite{Sidoli2009:sfxts_paperIV}); 
(e) IGR~J17544$-$2619  (2010-03-04, \cite{Romano2011:sfxts_paperVII}); 
(f) SAX~J1818.6$-$1703 (2009-05-06, \cite{Sidoli2009:sfxts_sax1818}); 
(g) AX~J1841.0$-$0536  (2010-06-05, \cite{Romano2011:sfxts_paperVII}). 
Adapted from \cite{Romano2011:sfxts_paperVII}.  \\

{\bf Right: (h)} 
\sw/XRT 0.2--10\,keV light curve of IGR~J18483$-$0311 during 
our monitoring program along {\bf one orbital period} (2009 June 11 to 2009 July 9)
at a binning of at least 20 counts bin$^{-1}$. 
 Data from \cite{Romano2010:sfxts_18483}.  \\

{\bf Right (i--l):} \sw/XRT 0.2--10\,keV light curves of our sample 
for a {\bf long-term monitoring program} (Table~\ref{texas10:tab:campaign}, 
2007 October 26 to 2009 November 3).  
Each point refers to the average flux observed
during each observation performed with XRT, except for outbursts 
where the data were binned to 
include at least 20 counts bin$^{-1}$ to best represent the 
dynamical range. Downward-pointing arrows are 3-$\sigma$ upper limits, 
upward pointing arrows mark either outbursts that  
XRT could not observe because the source was Sun-constrained,
or BAT Transient Monitor bright flares. 
AX~J1841.0$-$0536 was only observed during the first year. 
Data from \cite{Romano2009:sfxts_paperV,Romano2011:sfxts_paperVI}. 
}
\label{texas10:fig:alllcvs}
\end{figure}

The IGR~J11215$-$5952 results were initially met with some skepticism, and 
IGR~J11215$-$5952 was often considered an `odd ball' in the SFXT court. Now, however,
we have collected high-sensitivity soft X--ray light curves for the great majority of the 
SFXT sample, which we present in Figure~\ref{texas10:fig:alllcvs} (a-g). 
Furthermore, in Figure~\ref{texas10:fig:alllcvs} (h) we show the light curve of IGR~J18483$-$0311  \cite{Romano2010:sfxts_18483}, 
one of the three SFXTs for which both orbital and spin (21\,s, \cite{Levine2006:igr18483}) 
periods are known. These observations 
were obtained as a 28\,d monitoring program to study, 
for the first time, a SFXT along an entire orbital period (18.52\,d, \cite{Sguera2007}). 
We note that while obvious individual differences in the light curves are present, 
so are similarities, such as outburst length in excess of hours, a multiple peaked structure, 
and a very large dynamic range. Furthermore, there is ubiquitous variability on several 
timescales. 

Figure~\ref{texas10:fig:alllcvs} (i--l) shows the XRT light curves of the 
two-year campaign \cite{Romano2009:sfxts_paperV,Romano2011:sfxts_paperVI}. 
Several characteristic can be noted. \\ 
{\it i)} The four sources present a dynamical range of $\sim3$--4
orders of magnitude; AX~J1841.0$-$0536 went into outburst after the end of the campaign,
on 2010 Jun 5 and reached a dynamical range of $\sim 1600$ 
(see Fig~1 of \cite{Romano2011:sfxts_paperVII}). \\
{\it ii)} The long-term behaviour of SFXTs is not quiescence 
(which should be characterized by a soft spectrum and an X--ray 
luminosity of $\sim 10^{32}$\,erg\,s$^{-1}$) but an intermediate state 
of accretion with an average X-ray luminosity of $10^{33}$--$10^{34}$\,erg\,s$^{-1}$,
and a power law photon index $\Gamma=1$--2). \\
{\it iii)} We observe variability at all timescales we can probe. 
Superimposed on the day-to-day variability (Figure~\ref{texas10:fig:alllcvs} (i--l)),  
is intra-day flaring that involves flux variations up to one order of magnitude
(see, e.g.\ Figure~\ref{texas10:fig:alllcvs} a--g, and h).  
We can identify flares down to a count rate in the order of 0.1\,counts s$^{-1}$ 
($L\sim2$--$6\times10^{34}$ erg s$^{-1}$) within a snapshot of about 1\,ks. 
This short time scale variability cannot be accounted for by accretion from a 
homogeneous wind, but it can naturally  be explained 
by the accretion of single clumps in the donor wind. 
Assuming that each of these short flares corresponds to the accretion of a single clump 
onto the neutron star, then we can estimate its mass as 
$M_{\rm cl}= 7.5\times 10^{21} \,\, (L_{\rm X, 36}) (t_{\rm fl, 3{\rm ks}})^{3}$ g (\cite{Walter2007}) 
where $L_{\rm X, 36}$ is the average X-ray luminosity in units of $10^{36}$ erg s$^{-1}$,
$t_{\rm fl, 3{\rm ks}}$ is the duration of the flares in units of 3\,ks. 
We thus obtain $M_{\rm cl} \sim 0.3$--$2\times10^{19}$ g, which are about those expected 
(\cite{Walter2007}) to be responsible of short flares, below the \inte{} 
detection threshold and which, if frequent enough, may significantly contribute to the 
mass-loss rate. 

Given the structure of the observing plan during our campaign, 
we can realistically consider our monitoring as a 
casual sampling of the light curve at a resolution of about 4\,d. 
Therefore, we can calculate the percentage of time each source spent in each relative 
flux state. 
To this end, we selected three states, namely, the BAT-detected outbursts,
the intermediate states (observations yielding a firm detection excluding outburst ones),
and `non detections' (detections with a significance below 3$\sigma$). 
Only observations with an exposure in excess of 900\,s were considered 
(corresponding to flux limits $F_{\rm 2-10\,keV}^{\rm lim}\sim(1$--$3)\times 10^{-12}$\,erg\,cm$^{-2}$\,s$^{-1}$, 
depending on the source, \cite{Romano2009:sfxts_paperV}). 

The 4 SFXTs we monitored spend between 3 and 5\,\% of the time in {\it bright outbursts}.

The most {\it most probable flux level} at which a random observation
will find these sources, when detected, is $\sim 1$--$2\times10^{-11}$ erg cm$^{-2}$ s$^{-1}$ 
(unabsorbed 2--10\,keV), corresponding to luminosities of 
 a few $10^{33}$ to a few $10^{34}$ erg s$^{-1}$. These values are  
based (\cite{Romano2011:sfxts_paperVI}) on the distributions of the observed 
count rates after removal of the observations where a detection was not achieved
and are  two orders of magnitude lower than the bright outbursts, and two orders of magnitude 
higher than the quiescent state. 

Finally, the time each source spends {\it undetected} down to a $F_{\rm 2-10\,keV}^{\rm lim}=
1$--3$\times10^{-12}$\,erg\,cm$^{-2}$\,s$^{-1}$, which we defined as {\bf inactivity duty cycle} (IDC, 
\cite{Romano2009:sfxts_paperV}) is 
${\rm IDC}= \Delta T_{\Sigma} / [\Delta T_{\rm tot} \, (1-P_{\rm short}) ] \, , $  where  
$\Delta T_{\Sigma}$ is sum of the exposures accumulated in all observations (with exposure $>900$\,s)  
where only a 3-$\sigma$ upper limit was achieved 
   (Table~\ref{texas10:tab:campaign}, col.\ 5), 
$\Delta T_{\rm tot}$ is the total exposure accumulated (Table~\ref{texas10:tab:campaign}, col.\ 4), and 
$P_{\rm short}$ is the percentage of time lost to short observations 
   (exposure $<900$\,s, Table~\ref{texas10:tab:campaign}, col.\ 6). 
The cumulative count rate for each object is also reported 
Table~\ref{texas10:tab:campaign} (col.\ 8).
We find that ${\rm IDC} = (19, 28, 39, 55) \pm 5$\,\%, 
for IGR~J16479$-$4514, AX~J1841.0$-$0536, XTE~J1739--302,  IGR~J17544$-$2619, respectively
(Table~\ref{texas10:tab:campaign}, col.\ 7). 

\section{Conclusions}

Thanks to \sw\ we can now investigate the properties of SFTXs on 
several timescales (from minutes to days, from weeks to years)
and in several intensity states (bright flares,
intermediate intensity states, and down to almost quiescence); 
we can also perform broad-band spectroscopy of outbursts. 
Despite individual differences, common X--ray characteristics of this class 
are now well defined, such as outburst lengths well in excess of hours, with a
multiple peaked structure, and a high dynamic range of   
up to $\sim4$ orders of magnitude.

\acknowledgments
PR would like to thank the P4 Session Chairs, I.\ Grenier and M.\ Tavani,
for this chance to give an {\it impromptu} talk on the {\it Swift} results on SFXTs.
We acknowledge financial contribution from the agreement ASI-INAF I/009/10/0. 
This work was supported at PSU by NASA contract NAS5-00136.

\end{document}